
%
%
%

\font\eightrm=cmr8

\font\twelverm=cmr12
\font\lbf=cmbx10
\def\section#1{\par\vskip1.0\baselineskip
  \noindent{\lbf #1\hfil}\nobreak\vskip0.5\baselineskip\nobreak}

\font\caps=cmcsc10
\def\simle{\rlap{\raise 2pt \hbox{$<$}}{\lower 2pt \hbox{$\sim$}}}
\def\Hi{\hbox{H {\caps i}}}
\def\degree{^\circ}
\def\etal{{\it et al.}}

\null\vskip 0.8truecm\nobreak
\noindent{\twelverm Dark matter annihilations in the Large Magellanic 
Cloud\footnote{*}{\eightrm To appear in the proceedings of TAUP93, Gran Sasso 
Lab., Italy, 19-23 September 1993.}}
\vskip\baselineskip
\noindent Paolo\ Gondolo\footnote{$^{\rm a}$}{\eightrm Present address: 
LPTHE, Universit\'e Paris VII, Tour 24--14, 5 \'etage, 
2 Place Jussieu, 75251 Paris C\'edex 05, France; \hfill\break
e-mail: gondolo@lpthe.jussieu.fr}
\vskip\baselineskip
\noindent 
Department of Radiation Sciences, Uppsala University,
P.O.\ Box~535, 75121 Uppsala, Sweden
\vskip\baselineskip
\nobreak
{\parindent = 2em 
I examine the possibility of detecting high energy $\gamma$-rays from
non-baryonic dark matter
annihilations in the central region of the Large Magellanic Cloud.}
\nobreak
\section{1. LMC ROTATION CURVE}

The study of the kinematics of globular clusters in the outer regions of the
Large Magellanic Cloud by Schommer \etal\ [1] suggested a flat rotation
curve, perhaps out to 15 kpc: the LMC might have a dark matter halo.  Here,
from a composite \Hi\ and star cluster rotation curve (fig.~1), I want to
estimate the parameters of an isothermal halo added to a `maximum disk.'

Luks and Rohlfs [2], in the analysis of their extended 21-cm line survey of
the LMC, showed that the \Hi\ kinematics can be modeled by that of a flat disk
in differential rotation. The rotation curve they obtained considering data
within a $\pm 5\degree$ sector of the major axis is shown by crosses in
fig.~1. The geometrical and kinematical parameters they assumed
for the LMC are as follows: distance 50 kpc; inclination
33$\degree$; position angle of the major axis 162$\degree$;
\Hi\ kinematic center $ ( \alpha_0 , \delta_0 ) $
= $ (5^{\rm h} 12^{\rm m} 48^{\rm s}, -69\degree 03'.6) $; 
heliocentric velocity of the \Hi\ kinematic center 282.4 km/s radially
and 483 km/s tangentially towards 87$\degree$ position angle.

Using for consistency the same geometrical and kinematical parameters as
in ref.~[2], I have redetermined the star cluster rotation curve
in Schommer \etal\ 
[1], reading the heliocentric velocities and
the positions of the 83 star clusters from their tables 1 and 2. To render
the cluster rotation curve symmetric, the radial systemic velocity was
adjusted by 6.9 km/s. The points
shown as circles in fig.~1 were obtained using the clusters
lying within $\pm 30\degree$ of the major axis and beyond 3 kpc of the center.
The velocities have been binned in 1 kpc bins. The
error bars represent the standard deviation of the mean of velocities,
while circles without error bars are single objects. 

The rotation curve  cannot be firmly determined from these
data points. Nevertheless let us try to compare models with and without a
dark matter halo.

The most recent surface photometry of the LMC is provided by the CCD
observations of Bothun and Thompson [3] with a camera set up in a parking
lot.  An exponential disk with central brightness $ \mu_0 = 21.17 $
mag/arcsec$^2$ and scale length 1.$\degree$677 (1.46 kpc) fits the 
surface-brightness profile well in the innermost 3$\degree$ (2.6 kpc). (Values
in parentheses are for the assumed LMC distance of 50 kpc.) I use the
parameters of this fit plus a constant mass-to-luminosity ratio to
derive the rotation curve due to the exponential disk. Since signs of
truncation are present beyond 2.6 kpc, this choice somewhat overestimates the
disk contribution in the outer regions. The dashed curve in
fig.~1 corresponds to a disk M/L = 3.3 in solar units. Higher
mass-to-luminosity ratios would give curves overshooting the radio data points.

I now add a halo with canonical density profile $ \rho(r) = \rho_0 a^2 / ( r^2
+ a^2 ) , $ where $a$ is the core radius and $\rho_0$ is the central dark
matter density. Imposing that the disk+halo rotation curve passes close 
to the outermost points, the combination $ \rho_0 a^2,
$ which corresponds to the asymptotic rotation velocity, is constrained to be
$\approx$ 5 10$^{-24}$ g/cm$^3$ $\cdot$ kpc$^2$.
Unfortunately, a determination of the two halo parameters separately cannot be
achieved with the present data. The core radius is presumably $\ll 10$ kpc
and will be kept as a
parameter. For sake of illustration, a rotation curve with $a=1$ kpc
and $\rho_0 = $ 5  10$^{-24}$ g/cm$^3$ is plotted as a solid line in fig.~1.

\break
\section{2. GAMMA-RAYS FROM WIMP ANNIHILATION}

In this section, I entertain the possibility that weakly interacting massive
particles (WIMPs) constitute an LMC dark matter halo with $ a \ll 10$ kpc. An
indirect signature of their presence would be the $\gamma$-ray flux produced
after their self-annihilation.

The general expression of the $\gamma$-ray brightness from WIMP annihilation
in a canonical halo can be found in refs.~[4,5].  For a typical annihilation
rate $ \sigma v $ of 10$^{-27} $ cm$^3$/s and the above-mentioned LMC halo
parameters, the WIMP-generated
$\gamma$-ray brightness from the LMC center amounts to $$ I_\gamma = 3.0 \,
10^{-7} \, N_\gamma(E_\gamma) \, \hbox{\rm cm$^{-2}$ s$^{-1}$ sr$^{-1}$} 
 \, m_{10}^{-1} \, a_{\rm kpc}^{-3}
, $$ where $ N_\gamma(E_\gamma) $ is the number of photons of energy
$E_\gamma$ generated per annihilation, $a_{\rm kpc}$ is the core radius in kpc,
and $ m _{10} $ is the WIMP mass in units of 10 GeV. This brightness is $ \sim
9/a_{\rm kpc}^3 $ times bigger than the $\gamma$-ray flux used by Bengtsson \etal\ [5] in
their discussion of $\gamma$-ray signals from annihilation of galactic WIMPs
in the direction of the galactic pole.  By adopting their Lund Monte Carlo
estimates of $ N_\gamma(E_\gamma) $, I can simply rescale their $\gamma$-ray spectra up
by a factor of $ 9/a_{\rm kpc}^3 $.  

A possible foreground to the LMC WIMP-annihilation signal is 
the $\gamma$-ray flux from annihilation of galactic WIMPs in the direction of
the LMC. Using
the galactic parameters of ref.~[5], the latter is a factor 0.043 $a_{\rm
kpc}^3$ lower than
the former, and will be neglected in the following.

For interesting WIMP masses, 10 -- 1000 GeV, the background to the
annihilation signal is simply not known. At lower energies, $\gamma$-ray
emission from the LMC has been detected by EGRET aboard the Compton Gamma Ray
Observatory [6]. After subtraction of the galactic and extragalactic
components from the contour map in ref.~[6], the LMC emission in the central
region is $\sim$ 4 10$^{-6}$ cm$^{-2}$ s$^{-1}$ sr$^{-1}$  ($E_\gamma>100$
MeV).  This is at the level predicted for $\gamma$-ray production by
collisions of cosmic rays with the interstellar matter in the LMC. Using the
expected spectral index to extrapolate to higher energies, this 
background
is a factor of $\sim 4$ lower than the background adopted as a comparison
in ref.~[5]. If non-LMC emission would be included in the background, the last factor
would reduce to $\sim 2$.

The total gain in signal-to-background ratio with respect to the galactic case
examined in ref.~[5] is therefore of $\sim 30/a_{\rm kpc}^3$.
 This allows the signal-to-background
ratio to exceed unity in the most favorable cases if $a \simle $ few kpc.
 Since the background is
expected to have spectral index -1.7 above 300 MeV, the signal-to-noise ratio
would be almost independent of the WIMP mass if the analysis would be carried
out on the $\gamma$-ray spectrum itself. Knowledge of the latter is presently
prevented by the limited number of photons detected by EGRET.

\section{3. CONCLUSIONS}

If the Large Magellanic Cloud has a non-baryonic dark matter halo with
core radius $a \simle$ few kpc, it would be more
promising to search for $\gamma$-rays
 from dark matter annihilations in the
central regions of the LMC than in our galaxy. Detection of
such annihilation signals would be indirect evidence for the presence of
exotic matter in the LMC.  Further observations of the $\gamma$-ray emission
from the LMC central region could therefore have profound cosmological
implications.  

\vskip1.0\baselineskip

I kindly thank J Kaplan, M Persic, P Salucci and B Westerlund. Special thanks
go
to the organizers of TAUP93 for allowing
this contribution to appear in printing in spite of my forced absence from the
Conference.

\section{REFERENCES}
\vbox{\leftskip = 0.8em \parindent = 1em
\item{[1]} R A Schommer, E W Olszewski, N B Suntzeff and H C Harris, {\it
Astron J\/} 103 (1992) 447.
\item{[2]} Th Luks and K Rohlfs, {\it Astron Astrophys\/} 263 (1992) 41.
\item{[3]} G D Bothun and I B Thompson, {\it Astron J\/} 96 (1988) 877.
\item{[4]} J E Gunn \etal, {\it Astrophys J\/} 223 (1978) 1015;
 M Turner, {\it Phys Rev\/} D34 (1986) 1921.
\item{[5]} H-U Bengtsson, P Salati and J Silk, {\it Nucl Phys\/} B346 (1990)
129.
\item{[6]} P Sreekumar \etal, {\it Ap J\/} 400 (1992) L67.
}

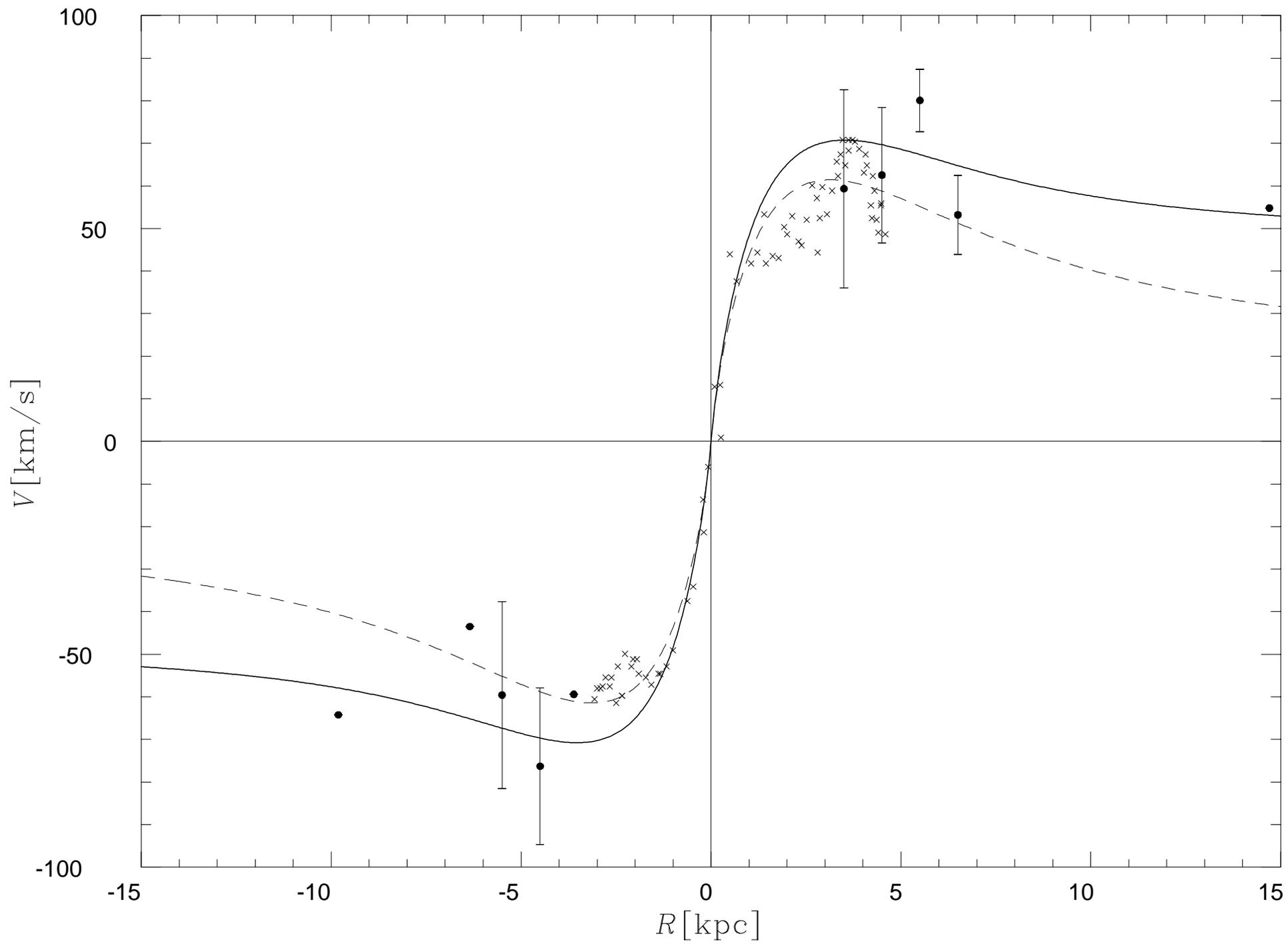
\bye